\documentclass{INTERSPEECH2023}
\usepackage{multirow}
\usepackage{pifont}
\usepackage{ragged2e}
\usepackage{makecell}
\interspeechcameraready

\title{Improving Speaker Verification with Self-Pretrained Transformer Models}
\name{Junyi Peng$^{1}$, Old\v{r}ich Plchot$^{1}$, Themos Stafylakis$^{2}$, Ladislav Mo\v{s}ner$^{1}$,   Luká\v{s} Burget$^{1}$, Jan \v{C}ernocký$^{1}$}
\address{
$^1$Brno University of Technology, Faculty of Information Technology, Speech@FIT, Czechia \\
$^2$Omilia - Conversational Intelligence, Athens, Greece
}
\email{}

\begin{document}
\maketitle
\begin{abstract}
Recently, fine-tuning large pre-trained Transformer models using downstream datasets has received a rising interest. Despite their success, it is still challenging to disentangle the benefits of large-scale datasets and Transformer structures from the limitations of the pre-training. In this paper, we introduce a hierarchical training approach, named self-pretraining, in which Transformer models are pretrained and finetuned on the same dataset. Three pre-trained models including HuBERT, Conformer and WavLM are evaluated on four different speaker verification datasets with varying sizes. Our experiments show that these self-pretrained models achieve competitive performance on downstream speaker verification tasks with only one-third of the data compared to Librispeech pretraining, such as VoxCeleb1 and CNCeleb1. Furthermore, when pre-training only on the VoxCeleb2-dev, the Conformer model outperforms the one pre-trained on 94k hours of data using the same fine-tuning settings.
\end{abstract}

\section{Introduction}
In recent years, deep neural networks based speaker recognition systems, such as x-vector \cite{snyder2018x}, ECAPA-TDNN \cite{desplanques2020ecapa}, and ResNet \cite{zhou2021resnext}, have achieved state-of-the-art performance on various speaker verification tasks. These models are typically trained in a supervised manner from scratch on a large-scale dataset like VoxCeleb \cite{nagrani2017voxceleb, chung2018voxceleb2} and CNCeleb \cite{li2020cn}, which contain millions of speech segments from thousands of speakers. However, due to their heavy parameterization, with tens of millions of parameters, optimizing these models can be challenging in data-restricted scenarios such as a completely new channel, language, or both in a low-resource domain.

A promising solution is to leverage general-purpose pre-trained Transformer models such as Wav2Vec \cite{ baevski2020Wav2Vec}, HuBERT \cite{hsu2021hubert}, and WavLM \cite{chen2022wavlm}, which are pre-trained on the large unlabelled test using self-supervised objective, like masked frame prediction, to learn general acoustic representations.
To adapt these pre-trained models to specific downstream tasks, a common approach is to perform fine-tuning of the entire pre-trained model with a task-oriented back-end using labeled downstream datasets. For example, in \cite{fan2020exploring}, an average pooling and a fully-connected layer were cascaded to the top of the Wav2vec 2.0 model to extract speaker and language embedding. A stronger performance was achieved on the speaker verification task using back-end, i.e. ECAPA-TDNN, in \cite{chen2022large}, which fed the frame-by-frame input to the back-end as a weighted combination of the outputs of the individual layers of a pre-trained Transformer model. To shorten the training time, a more lightweight back-end was developed in \cite{peng2022attention}, which consists of an attention layer and a linear layer to extract speaker representations. These simple approaches have led to impressive results in the fields of language recognition \cite{lee20232022} and speaker verification tasks. Despite their success, it remains challenging to disentangle the advantages offered by large-scale datasets and Transformer structures  from the limitations of the pre-training paradigm \cite{el2021large}. 

The challenges can be summarized as three-fold considering the mask prediction objective is the optimization target for Transformer models. (1) To what extent large-scale datasets are crucial for performance improvements? Particularly, in cases where sufficient downstream data is available, is there still a need to employ pre-trained models? For instance, the HuBERT base is pretrained on 960-hour Librispeech, while the fine-tuning dataset, VoxCeleb2, comprises over 2,000 hours of speech. (2) How does domain shift affect pre-trained models?  Currently, most pre-trained models are trained on English read-aloud data, raising questions about their effectiveness when fine-tuned in non-English language and under in-the-wild conditions. For example, a pre-trained WavLM Base model may not perform well when fine-tuned on a Chinese speaker recognition corpus \cite{li2020cn, fan2020cn}. The phonetic and linguistic properties of different languages can significantly vary, and models may not capture these nuances effectively. (3)  What is the impact of the Transformer structure on downstream speaker verification tasks? The widely-used HuBERT structure is inherited from BERT \cite{devlin2018bert}, which is originally designed for natural language processing tasks. It still remains unclear whether this structure is optimal for downstream speaker verification tasks. In addition, recent studies have shown performance improvement in speech recognition using alternative architectures such as Conformer \cite{gulati2020conformer} and WavLM  \cite{chen2022wavlm}. 

\begin{figure}
% \begin{minipage}[t]{.49\linewidth}
%   \centering
%   \centerline{\includegraphics[width=\linewidth]{Fig/Fig0.png}}
%   \centerline{(a) Generalist Pretraining}\medskip
% \end{minipage}
% \hfill
% \begin{minipage}[t]{0.49\linewidth}
%   \centering
%   \centerline{\includegraphics[width=\linewidth]{Fig/Fig1.png}}
%   \centerline{(b) Self Pretraining}\medskip
% \end{minipage}
%
\centering
\includegraphics[width=\linewidth]{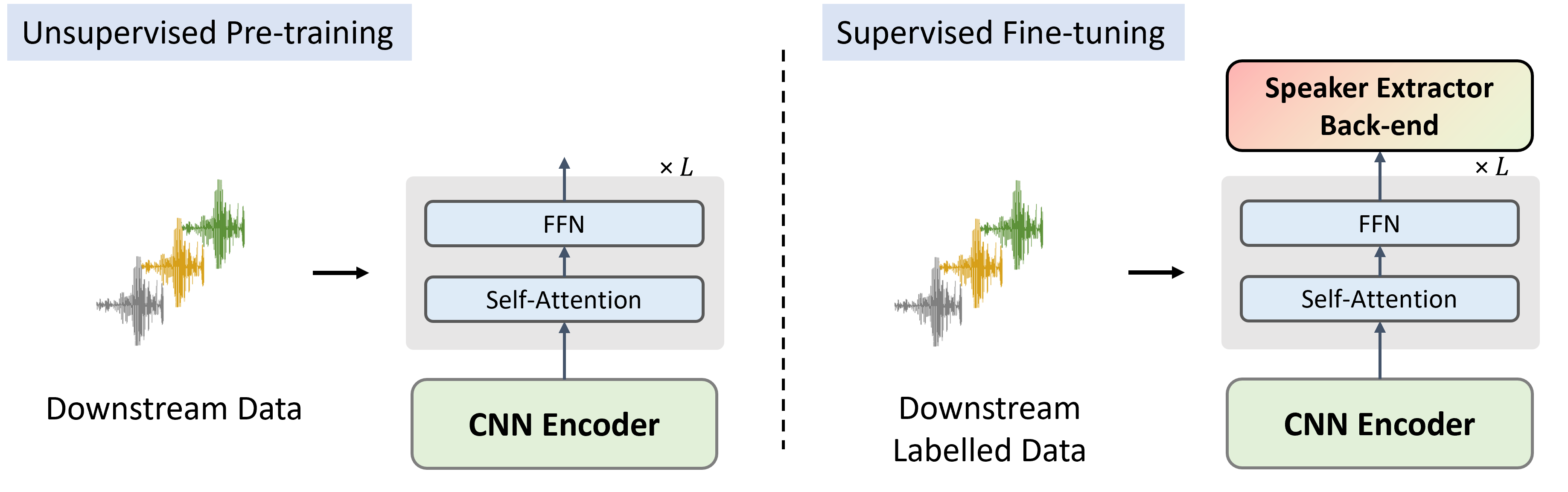}

\caption{In contrast to generalist pretraining, where a pre-trained model is first trained on a large-scale corpus (e.g., Librispeech) before fine-tuning on downstream task, self-pretraining utilizes the model pretrained on the fine-tuning dataset itself. }
\label{fig:sys}
\end{figure}

To tackle these challenges, in this paper, we explore a hierarchical training paradigm, named \emph{self-pretraining}, which aims to reduce the domain shift between pre-training and fine-tuning stages and improve the discriminability of extracted speaker representations. Specifically, as shown in Fig \ref{fig:sys}, we construct the training process based on the following steps: firstly, the Transformer model is directly trained from scratch on the downstream datasets with the HuBERT-style mask prediction loss following the self-supervised manner. Then, a learnable task-oriented back-end is introduced to cascade the pre-trained Transformer model to predict the speaker identity based on a given utterance while the parameters of the pre-trained Transformer model are frozen.  Finally, the entire Transformer model is jointly fine-tuned with the back-end to further boost the performance.
Overall, our contributions are the following:
\begin{itemize}
    \item We demonstrate the effectiveness of self-pretraining across four different pre-training/fine-tuning datasets, with two cascaded back-end structures (ECAPA-TDNN and Multi-head Factorized Attentive Pooling (MHFA) \cite{peng2022attention}) for SV tasks.
    \item A detailed analysis of cross-language performance of models pretrained on one language and then fine-tuned on another language for a SV task.
    \item We conduct a comprehensive study on the SV performance of pre-trained models with different architectures including HuBERT, Conformer and WavLM (Transformer with gated position encoding and data augmentation).
    \item Extensive experiments on both VoxCeleb and CNCeleb corpus show that our methods obtain a significant improvement over Librispeech pre-trained systems and outperform the existing SV systems. 
\end{itemize}

\section{Self-pretraining for Speaker Verification}
\label{sec:RW}
In this section, we formalize each of the self-pretraining components in detail, as illustrated in Fig \ref{fig:sys}. Our self-pretraining approach explores the potential of pre-trained models by utilizing the same dataset for both pre-training and fine-tuning while minimizing domain-shift between two datasets. Furthermore, the model is expected to learn in-domain representations that effectively capture the phonetic and acoustic characteristics of speech, as well as contextual information that enables it to predict masked frames. These learned representations can potentially be beneficial in the field of speaker verification, where the model is supposed to accurately distinguish between subtle speech signal variations to identify the speaker. Although similar observations have been reported in computer vision \cite{el2021large} and natural language processing \cite{krishna2022downstream}, it is essential to validate these findings in the field of speaker verification, where the use of pretrained models has attracted increasing attention.

\subsection{Unsupervised Pre-training}
In this study, we utilize the learned representations from three different architectures, including HuBERT, Conformer, and WavLM, respectively, as input features to extract speaker embeddings. These models take raw waveforms as input, and the backbone consists mainly of a convolutional neural network (CNN) encoder and a Transformer encoder \cite{vaswani2017attention}. 

During pre-training, the model consumes masked frame-level features to predict a predetermined discrete target. This masked speech prediction objective is applied only to the masked frame-level features, with the model expected to correctly infer the targets of masked frames through the remaining unmasked ones. Hence, the model is enforced to learn acoustic and phonetic information over continuous input speech.

We follow the iterative re-clustering and re-training approach described in \cite{hsu2021hubert}. In the initial iteration, targets are assigned by clustering the MFCC features of the training data. In subsequent iterations, a new set of training targets is generated by clustering the latent representations produced by the first iteration trained model.

\subsection{Supervised Fine-tuning}

Transformer models trained on thousands of hours of speech data have been shown to effectively represent the structure of speech and generalize well to various downstream tasks \cite{wang2021unispeech, chen2022unispeech}. However, recent studies have suggested that the advantage of fine-tuning pre-trained models over deep speaker extractors trained from scratch is insignificant or even non-existent, based on analysis of the last layer's representations \cite{fan2020exploring, wang2021fine, tak2022automatic}. A potential explanation is that the speech prediction objective of the pre-training stage encourages the model to discover and internally represent various acoustic units that naturally correspond to context-dependent phones in the last layers \cite{laperriere2022use}. As a result, the top layers, which are closer to the objective of the pre-training stage, are typically most beneficial for automatic speech recognition. In contrast, speaker verification tasks mainly rely on low- and mid-level features that carry most of the information about speaker identity. Thus, attempting to obtain the speaker representations from the last Transformer layer's output may be sub-optimal for speaker verification.

Based on the aforementioned, to fully utilize the hierarchical representations, we follow the method introduced in \cite{yang2021superb} and we assign a set of learnable weights $\mathbf{w} = \{w_{l}\}_{l=0}^{L}$ to the layer-wise outputs as:
\begin{equation}
\label{eq1}
\begin{split}
\mathbf{y} = f_{b} \left( \sum_{l=0}^{L}w_{l}\mathbf{H}_{l}; \boldsymbol{\Theta}_{b} \right),
\end{split}
\end{equation}
where $l$-th layer's outputs are $\mathbf{H}_l \in \mathbb{R}^{T\times D}$ ($\mathbf{H}_0$ denotes the CNN outputs), $L$ denotes the total number of Transformer layers, $T$ is the length of frames, $D$ is the feature dimension, $\boldsymbol{\Theta}_{b}$ are the parameters of the backend, and $\mathbf{y}$ is the extracted speaker embedding. Regarding the back-end, in this study, we utilize two different structures: ECAPA-TDNN and MHFA. In detail, for ECAPA-TDNN we replace the original FBank features inputs with hierarchical representations obtained from the pre-trained Transformer model. For MHFA, we modify eq. (\ref{eq1}) so that \emph{two sets of layer-wise weights} ($\mathbf{w}^K$ and $\mathbf{w}^V$) are employed to generate attention maps and compressed features, respectively \cite{peng2022attention}. Then, the speaker embedding $\mathbf{y}$ is formed by aggregating over frames and projecting the vector to a lower-dimensional space using a linear layer.

\begin{table*}[t] 
  \caption{Performance comparison of HuBERT Base models pretrained on various datasets, including Librispeech, VoxCeleb and CNCeleb. \textbf{RandomInit} indicates both Transformer and back-end are jointly trained from scratch on a fine-tuning dataset. \textbf{Librispeech} denotes the models are pre-trained on the Librispeech dataset from Huggingface. \textbf{Self-Pretrain} means model pre-training and finetuning on the same dataset. \textbf{Frozen} suggests the pretrained models are kept frozen during the fine-tuning, while \textbf{Learnable} denotes the pre-trained model jointly optimized with the back-end. It is noted that here we use \textbf{ECAPA-TDNN} as the back-end for all systems.}
                    \vspace{-0.3cm}
  \label{tab:1}
  \centering
    \scalebox{1}{
    \begin{tabular}{c|c|c|c|cc|cc}  
    \hline
    \multicolumn{1}{c|}{\multirow{2}{*}{Fine-Tuning Dataset}} & \multicolumn{1}{c|}{\multirow{2}{*}{Hours(hr)}}  & \multicolumn{1}{c|}{\multirow{2}{*}{Test Dataset}}  & \multicolumn{1}{c|}{\multirow{2}{*}{RandomInit}} & \multicolumn{2}{c|}{Librispeech (960 hr)} & \multicolumn{2}{c}{Self-pretraining} \cr \cline{5-8} &&&& Frozen & Learnable & Frozen & Learnable\\
    \hline
    \hline
    VoxCeleb1 & 300  & VoxCeleb1-O & 10.14 &2.43 & 2.22 & 2.14 & 1.91\\
    VoxCeleb2-dev & 2000  & VoxCeleb1-O & 1.88 &1.36 & 1.18 & 1.13 & 1.04\\
    CNCeleb1 & 270  & CNCeleb-E & 21.08 & 14.64 &  11.99 & 11.90 & 10.86\\
    CNCeleb1+2 & 1200  & CNCeleb-E & 15.33 & 10.65 & 9.70 & 8.80 &  8.89 \\
\hline
    \end{tabular}
                  \vspace{-0.3cm}
    }
\end{table*}

\begin{figure}[t]
    \centering
    \includegraphics[width=\linewidth]{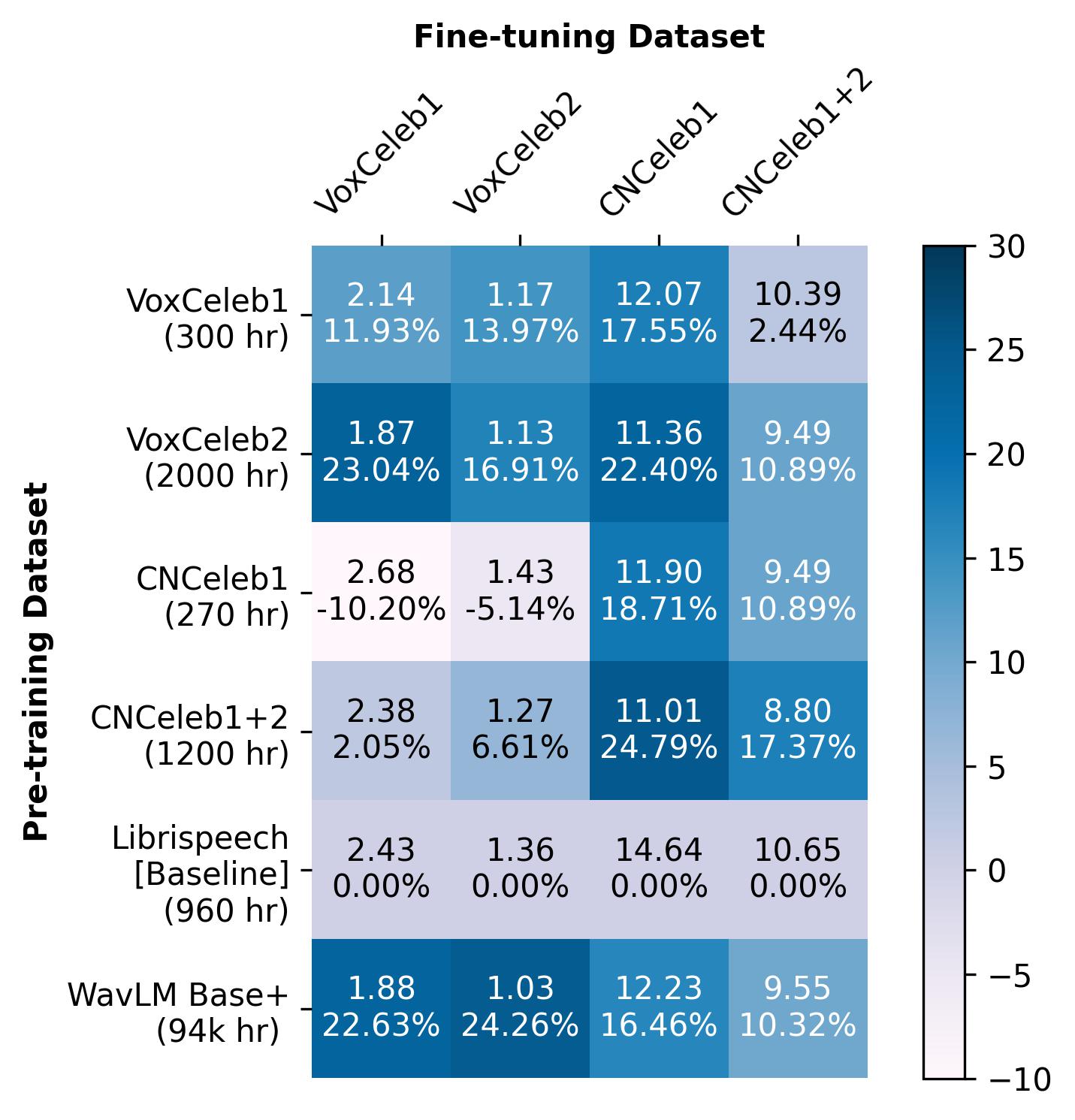}
                        \vspace{-0.3cm}
    \caption{Equal Error Rates (EERs) and improvements of models pretrained on various datasets, fine-tuned on downstream speaker verification tasks with different datasets. Each value represents the EER and relative improvement gains compared to the modeal readily available on HuggingFace pretrained on Librispeech. For systems fine-tuned on Voxceleb1 and Voxceleb2, the results are given on VoxCeleb1-O, while for systems fine-tuned on CNCeleb we report results on CNCeleb-E. During fine-tuning the pre-trained models are kept frozen. }
    \label{fig:exp2}
\vspace{-0.4cm}
\end{figure}

\begin{table}[t] 
  \caption{The performance of various Base models with different back-ends on VoxCeleb1-O. All models are self-pretrained on VoxCeleb2-dev. \textbf{WavLM} refers to a Transformer with gated relative position bias and simulated noisy/overlapped speech. }
                    \vspace{-0.3cm}
  \label{tab:2}
  \centering
    \scalebox{0.95}{
    \begin{tabular}{c|c|c|c}  
    \hline
Backbone/Param & Back-end & Finetune & EER (\%)\\
\hline
\multirow{3}{*}{\makecell{HuBERT\\(94.6 M)}} & ECAPA-TDNN & \ding{55} & 1.13\\
& ECAPA-TDNN & \ding{51} & 1.04\\
&  MHFA & \ding{55} & 1.38 \\
&  MHFA & \ding{51} & 0.88 \\
\hline
\multirow{3}{*}{\makecell{Conformer\\(172.2 M)}} & ECAPA-TDNN & \ding{55} & 1.17 \\
& ECAPA-TDNN & \ding{51} & 1.19 \\
&  MHFA & \ding{55} & 1.21 \\
&  MHFA & \ding{51} & 0.65 \\
\hline
\multirow{3}{*}{\makecell{WavLM\\(94.7 M)}} & ECAPA-TDNN & \ding{55} & 1.12\\
& ECAPA-TDNN & \ding{51} & 1.09\\

&  MHFA & \ding{55} & 1.40\\
&  MHFA & \ding{51} & 0.87\\
\hline
    \end{tabular}
    }
                  \vspace{-0.3cm}
\end{table}
% TODO The result for WavLM Base - MHFA will be updated in our final version.
\section{Experiments}
\subsection{Setup}
\textbf{Data-sets:} The SV performance is evaluated on the VoxCeleb \cite{nagrani2017voxceleb,chung2018voxceleb2} and CNCeleb \cite{fan2020cn,li2020cn} corpora, both are widely used text-independent speaker verification datasets. For VoxCeleb, the training set is VoxCeleb1, and the development set of VoxCeleb2, respectively. The performance is evaluated on \textit{VoxCeleb1-O}, \textit{VoxCeleb1-E}, and \textit{VoxCeleb1-H} trials. For CNCeleb, the model is trained on  different training datasets, namely, \textit{CNCeleb1} and \textit{CNCeleb1+2}, containing 800 and 2800 speakers, respectively. \textit{CNCeleb1+2} is a combination of CNCeleb1-dev and CNCeleb2. The evaluation part \textit{CNCeleb-E} contains 18,849 utterances from 200 speakers.

\noindent\textbf{Implementation details:}
In this work, we utilize three types of pre-trained BASE models: 1) The \emph{HuBERT} models, consisting of a CNN encoder and 12 layers of Transformer with 94M parameters. The dimension of the Transformer output $D_{hidden}$ is 768. 2) The \emph{Conformer} models, which include 12 layers combining both convolutional neural networks and transformers, result in 170M parameters. 3) The \emph{WavLM} models, a variant of the Transformer model that redesigns the position embedding and utilizes simulated noisy/overlapped speech to improve model's robustness. We then pre-train these models for 800k steps on VoxCeleb2-dev and CNCeleb1+2, and for 400k steps on VoxCeleb1 and CNCeleb1 using the labels generated by clustering the 6-th transformer layers outputs of 1st-iteration (250K steps) corresponding model following \emph{fairseq} implementation. The pre-training stage takes 3-4 days on 8 A100 GPUs. 

For fine-tuning configurations, the total number of heads in MHFA is set to 64, the channel of ECAPA-TDNN is 1024, and the extracted speaker embedding dimension is 256 for both back-ends. We use AAM-softmax \cite{deng2019arcface} in a fine-tuning stage with a margin of 0.2 and scaling of 30 for 10 epochs. To further boost performance, we adopt large margin tuning \cite{thienpondt2021idlab}. we input longer (5 seconds) waveforms and set the margin to 0.5 for additional 3 training epochs.
The initial learning rate is $5e^{-4}$ and decreases by $10\%$ per epoch for the Adam optimizer. All fine-tuning datasets are augmented by adding noise (MUSAN) and reverberation (RIR). Due to the GPU memory constraints, the mini-batch size of $100$ is chosen for Base model training. 
%Apart from joint fine-tuning, the pre-trained model is fixed. For faster computation and less GPU memory consumption, we use 16-bit float precision. For a fair comparison, we use cosine similarity. 

\noindent\textbf{Performance Metrics:}
Both equal error rate (EER) and minimum detection cost function (minDCF) are employed to measure the performances of speaker verification systems. The prior target probability $P_{\textit{tar}}$ is set to 0.01 or 0.05, for DCF1 and DCF5, respectively. $C_{\textit{fa}}$ and $C_{\textit{miss}}$ are set to 1.0.

\begin{table*}[t] 
  \caption{Results for speaker verification on the Voxceleb1-O data-set and extended VoxCeleb1-E and VoxCeleb-H test sets. All models are fine-tuned on VoxCeleb2-dev. \textbf{Pre-train} denotes the pre-training dataset.}
                    \vspace{-0.3cm}
  \label{tab:3}
  \centering
    \scalebox{0.88}{
    \begin{tabular}{l|c|ccc|ccc|ccc}  
    \hline  
    \multicolumn{1}{c|}{\multirow{2}{*}{Model}}&\multicolumn{1}{c|}{\multirow{2}{*}{Pre-train}}&\multicolumn{3}{c|}{VoxCeleb1-O}&\multicolumn{3}{c}{VoxCeleb1-E}&\multicolumn{3}{|c}{VoxCeleb1-H}\cr\cline{3-11}  & &EER(\%)&DCF1 &DCF5&EER(\%)&DCF1&DCF5 &EER(\%)& DCF1&DCF5 \\
    \hline
    \hline
         ECAPA-TDNN \cite{kwon2021ins} & - &  0.90 & - & 0.081 & 1.11 & - & 0.077 & 2.32 & - & 0.155 \\
    WavLM Base Plus- ECAPA-TDNN \cite{chen2022wavlm} & Mix 94k hr & 0.98 & - & - & 1.06 & - & - & 2.21 & - & - \\
    WavLM Base Plus- MHFA \cite{peng2022attention}& Mix 94k hr & 0.66 & 0.074 & 0.045 & 0.89 & 0.097 & 0.056 & 1.90 & 0.190 & 0.119\\
        HuBERT Base - MHFA \cite{peng2022attention}& LS 960 hr & 0.92 & 0.091 & 0.059 & 1.19 & 0.136 & 0.078 & 2.52 & 0.252 & 0.159 \\
        \hline
        HuBERT Base - MHFA & Vox2-dev 2k hr &  0.88 & 0.097 & 0.056 & 1.06 & 0.118 & 0.068 & 2.11 & 0.223 & 0.136\\
        WavLM Base - MHFA & Vox2-dev 2k hr &  0.87 & 0.091 & 0.053 & 1.04 & 0.114 & 0.072 & 2.23 & 0.227 & 0.133\\
        Conformer - MHFA & Vox2-dev 2k hr & 0.65 & 0.063 & 0.045 & 0.93 & 0.100 & 0.058 & 1.86 & 0.193 & 0.117 \\

\hline
    \end{tabular}
    }
                      \vspace{-0.4cm}

\end{table*}

\begin{table}[t]
    \caption{Comparison of different models on CNCeleb-E test set. All models are trained on the CNCeleb1+2 dataset.}
    \centering
                          \vspace{-0.3cm}
    % \begin{tabular}{p{100 pt}<{\RaggedLeft}|p{40 pt}<{\centering}|p{40 pt}<{\centering}}
        \begin{tabular}{p{4cm}<{\RaggedRight}|p{1.2cm}<{\centering}|p{1.2cm}<{\centering}}

    \hline
         Model &  EER(\%) & DCF1 \\
         \hline
         \hline
         ResNet34-DTCF \cite{zhang2021duality} & 14.84 & 0.596 \\
         MBFA-MW \cite{qin2023multi} &  9.48 & 0.456 \\
         ECAPA-TDNN \cite{zeng2022attention} & 8.93 & 0.504 \\
         \hline
         Conformer - MHFA & 7.73 & 0.406 \\
          \hline
    \end{tabular}
    \vspace{-0.4cm}
    \label{tab:4}
\end{table}

\subsection{Comparison of Generalist and Self Pretraining}
To investigate the performance of self-pretraining and other pretraining techniques. For each dataset, we first pretrain a HuBERT Base Transformer model from scratch and then fine-tune it on the same training dataset with speaker labels in Table \ref{tab:1}. 

Our experimental results demonstrate that the self-pretrained models consistently outperform the fine-tuned Librispeech pretrained models obtained from Huggingface, even when the amount of training data is significantly lower than that of Librispeech. Since the self-pretraining models are fine-tuned on the same dataset used for pretraining, it is unlikely that the observed performance gains are due to transfer learning from pre-training on a different dataset. Instead, the improvements may be attributed to the hierarchical training approach, which enables the models to learn in-domain representations and capture relevant acoustic and phonetic features of the downstream dataset during pretraining. By fine-tuning on the same dataset, the cascaded back-end is able to learn more speaker-related information based on these robust features, resulting in better generalization and performance improvement. Furthermore, when we jointly train the Transformer model and the followed back-end from scratch on speaker recognition tasks, referred to as RandomInit, the worst performance is attained. It is finally observed that when the training dataset is insufficient, such as in the case of VoxCeleb1 and CNCeleb1, the performance degrades dramatically.

\subsection{Analysis of Cross-Language Fine-tuning}
We investigate the generalizability of pre-trained models across a range of SV tasks with different scales and languages, as illustrated in Fig \ref{fig:exp2}. In detail, we explore whether models pre-trained on a particular dataset are only useful for that specific dataset or whether they can be further applied to a broader range of conditions, such as cross-language scenarios. We first take four Transformer models pre-trained on VoxCeleb1, VoxCeleb2-dev, CNCeleb1 and CNCeleb1+2, respectively. Subsequently, we fine-tune the back-end of each pre-trained Transformer on all other datasets. For the approaches fine-tuned on VoxCeleb1 and VoxCeleb2-dev, the performance is evaluated on VoxCeleb1-O; while for CNCeleb1 and CNCeleb1+2, we evaluate their performance on CNCeleb-E. The relative performance improvements and corresponding EERs are shown as a heatmap in Fig \ref{fig:exp2}.

In most cases, the self-pretrained models perform better than the Librispeech pretrained models. Moreover, we observe that the performance improvement is more significant when there is a larger amount of data in a similar scenario during fine-tuning. For example, pre-training on VoxCeleb2-dev and fine-tuning on VoxCeleb1 leading to a significant performance boost. However, it is noted that the Transformer model pre-trained on CNCeleb1 performs worse with negative benefits. This might be because the limited amount of training data (200 hours) may not be sufficient to train a Transformer model with 96 million parameters effectively. Moreover, compared to VoxCeleb1 with a similar size, CNCeleb1 is collected with a variety of challenging acoustic and phonetic conditions, such as regional accents, background noise, and music interference, which can make the pre-training task more difficult.

\subsection{Ablation Studies on Architectures}
To demonstrate the effectiveness of self-pretraining, in Table \ref{tab:2}, we present an analysis of the impact of various backbone architectures, including the original Transformer model, Conformer model, and WavLM model, and two back-end models (i.e. ECAPA-TDNN and MHFA). 

By replacing the Transformer model with the Conformer model, a substantial improvement is observed, especially when using the MHFA back-end to joint fine-tuning with the entire pre-trained model. This suggests that the depthwise separable convolution module in the Conformer model is more effective in capturing local acoustic features compared to the original Transformer model. 

Additionally, the system using MHFA is able to extract more discriminative speaker representations, yielding better performance on the VoxCeleb1-O compared to the system using ECAPA-TDNN. This may be due to the fact that the pre-trained model has already learned rich and robust acoustic representations, and the lightweight back-end with fewer parameters gives the pre-trained model more flexibility, allowing it to better adapt to the speaker verification tasks. In contrast, a heavy, multi-layer back-end, designed to be a standalone network, cannot make full use of the backbone network and propagate gradients to finetune it. As Table \ref{tab:2} shows, the gains from unfreezing the backbone network are marginal when ECAPA-TDNN is used.

\subsection{Comparison with State-of-the-art SV systems}
We compare the proposed method with other state-of-the-art SV systems in Tables \ref{tab:3} and \ref{tab:4}. All pre-trained model-based approaches achieve remarkable performance compared to conventional SV systems trained from scratch, indicating that pre-trained models can effectively capture relevant acoustic and phonetic information for downstream tasks. Among the pre-trained models, our self-pretraining approach consistently outperforms the generalist one. These results show the effectiveness of self-pretraining in capturing speaker-specific information, resulting in more discriminative embeddings.

\section{Conclusion}
In this paper, we propose a hierarchical training approach, named self-pretraining, for pre-training Transformer models on downstream speaker verification tasks. We have shown that this approach can capture in-domain and speaker-specific information, resulting in more discriminative speaker embeddings. Our experiments show that self-pretraining outperforms generalist pretraining and achieves competitive performance with 94k hours of pretraining on downstream speaker verification tasks with solely 2k hours of downstream data. Finally, we showed that the recently introduced MHFA is superior to ECAPA-TDNN as a backend for Transformer models. 

\bibliographystyle{IEEEtran}
\bibliography{mybib}

\end{document}